\documentclass{ws-ijmpa}

\begin{document}

\begin{center} 
{\Large $R$-Parity in Supersymmetric Left-Right Models.}
\end{center}

\begin{center}
M. C. Rodriguez  \\
mcrodriguez@fisica.furg.br \\
{\it Funda\c c\~ao Universidade Federal do Rio Grande-FURG \\
Departamento de F\'\i sica \\
Av. It\'alia, km 8, Campus Carreiros \\
96201-900, Rio Grande, RS \\
Brazil}
\end{center}

\date{\today}

\begin{abstract}
On this article we show explicity that Supersymmetric Left-Right Models already satisfy the $R$-parity. They also respect $L$-parity and $B$-parity.

\end{abstract}

\section{Introduction}

Although the Standard Model (SM) gives very good results in
explaining the observed properties of the charged fermions, it is
unlikely to be the ultimate theory. It maintains the masslessness
of the neutrinos to all orders in perturbation theory, and even
after non-pertubative effects are included. The recent
groundbreaking discovery of nonzero neutrino masses and
oscillations \cite{superk} has put massive neutrinos as one of
evidences on physics beyond the SM.

The Super-Kamiokande experiments on the atmospheric neutrino
oscillations have indicated to the difference of the squared
masses and the mixing angle with fair accuracy
\cite{hanoiconf,fogli}
\begin{eqnarray}
\label{eqn::atm}
\Delta m^2_{\mathrm{atm}}  & = & 1.3 \div 3.0 \times 10^{-3}  {\rm eV ^2}, \\
\sin^2 2 \theta_{\mathrm{atm}}  & > & 0.9.
\end{eqnarray}
While, those from the combined fit of the solar and reactor
neutrino data point to
\begin{eqnarray}
\label{eqn::solar}
\Delta m^2_{\odot} ~ & = &
8.0^{+ 0.6}_{- 0.4} \times 10^{-5} ~ \rm{ eV^2}, \\
\tan ^2 \theta_{\odot} ~ & = & 0.45^{+0.09}_{-0.07}.
\end{eqnarray}
Since the data provide only the information about the differences
in $m_{\nu}^2$, the neutrino mass pattern can be either almost
degenerate or hierarchical. Among the hierarchical possibilities,
there are two types of normal and inverted hierarchies. In the
literature, most of the cases explore normal hierarchical one in
each. In this paper, we will mention on a supersymmetric model
which naturally gives rise to three pseudo-Dirac neutrinos with an
inverted hierarchical mass pattern.

The gauge symmetry of the SM as well as those of many extensional
models by themselves fix only the gauge bosons. The fermions and
Higgs contents have to be chosen somewhat arbitrarily. In the SM,
these choices are made in such a way that the neutrinos are massless
as mentioned. However, there are other choices based on the SM
symmetry that neutrinos become massive. We know these from the
popular seesaw \cite{seesaw} and radiative \cite{rad} models.

Certainly a very popular extension of the SM is the left-right 
symmetric theories \cite{ps74}, which attribute the 
observed parity asymmetry in the weak interactions to the spontaneous 
breakdown of Left-Right symmetry, i.e. generalized parity transformations.
Furthermore, Left-Right symmetry plays an important role in attempting to
understand the smallness of CP violation \cite{bt78}.

Apart from its original motivation of providing a dynamic
explanation for the parity violation observed in low-energy weak interactions,
this model differs from the SM in another important aspect; it explains the
observed lightness of neutrinos in a natural way and it can also solve the
strong CP problem.

On the technical side, the left-right symmetric model has a problem similar to
that in the SM: the masses of the fundamental Higgs scalars diverge
quadratically. As in the SM, the SUSYLR can be used to stabilize the scalar
masses and cure this hierarchy problem.

On the literature there are two different SUSYLR models. They differ in their
$SU(2)_{R}$ breaking fields: one uses $SU(2)_{R}$ triplets (SUSYLRT) and the
other $SU(2)_{R}$ doublets (SUSYLRD). Theoretical consequences of these models
can be found in various papers including \cite{susylr} and \cite{doublet}
respectively.
 
Another,
 maybe more important {\it raison d'etre} for supersymmetric Left-Right
 models is the fact that they lead naturally to R-parity conservation.
Namely, Left-Right models contain a $B-L$ gauge symmetry, which allows
for this possibility \cite{m86}. All that is  needed is that one uses
a version  of the theory that incorporates a see-saw mechanism \cite{seesaw1}
at the renormalizable level. More precisely, R-parity (which keeps
particles invariant, and changes the sign of  sparticles) can be
written as
\begin{equation}
R = (-1)^{3 (B - L) + 2 S}
\label{rparity}
\end{equation}
where $S$ is the spin of the particle. It can be shown that in these
kind of theories, invariance under $B-L$ implies R-parity conservation
\cite{m86}. It is just the goal of this article. We will show that 
both models, SUSYLRT and SUSYLRD, are invariant under $R$-Parity transformation. Before, 
we start it is useful to remember that the choice of the
triplets is preferable to doublets because in the first case we can
generate a large Majorana mass for the right-handed neutrinos \cite{mfrank}.

The goal of this article is to show this appealing characteristics of having 
automatic $R$-parity conservation in both models SUSYLRT and SUSYLRD.

\section{$R$-Symmetry}

It is important to note that the SM can explain the conservation
of lepton number ($L$) and of baryon number ($B$) without needing
to any discrete symmetry. However, this is not the case of
supersymmetric theories where only if interactions of conserving
both $L$ and $B$ are required, one has to impose one discrete
symmetry. 

The R-symmetry was introduced in 1975 by A. Salam and J. Strathdee
\cite{r1} and in an independent way by P. Fayet \cite{r2} to avoid
the interactions that violate either lepton number or baryon
number. There is very nice review about this subject in
Refs.\cite{barbier,moreau,hung}.

\subsection{Discrete R-Parity in SUSYLRT}

\begin{table}[t]
\center
\renewcommand{\arraystretch}{1.5}
\begin{tabular}
[c]{|l|cc|cc|}\hline
Superfield & Usual Particle & Spin & Superpartner & Spin\\\hline\hline
\quad$\hat{V}^{\prime}$ (U(1)) & $B_{m}$ & 1 & $\tilde{B}\,\,$ & $\frac{1}{2}
$\\
\quad$\hat{V}^{i}_{L}$ ($SU(2)_{L}$) & $W^{i}_{mL}$ & 1 & $\tilde{W}_{L}^{i}$
& $\frac{1}{2}$\\
\quad$\hat{V}^{i}_{R}$ ($SU(2)_{R}$) & $W^{i}_{mR}$ & 1 & $\tilde{W}_{R}^{i}$
& $\frac{1}{2}$\\
\quad$\hat{V}^{a}_{c} (SU(3))$ & $g^{a}_{m}$ & 1 & $\tilde{ g}^{a}$ &
$\frac{1}{2}$\\\hline
\quad$\hat{Q}_{i}\sim({\bf3},{\bf2},{\bf1},1/3)$ & $(u_{i},\,d_{i})_{iL}$ &
$\frac{1}{2}$ & $(\tilde{ u}_{iL},\,\tilde{ d}_{iL})$ & 0\\
\quad$\hat{Q}^{c}_{i}\sim({\bf3^{*}},{\bf1},{\bf2},-1/3)$ & $(d^{c}_{i}
,\,-u^{c}_{i})_{iL}$ & $\frac{1}{2}$ & $(\tilde{ d}^{c}_{iL},\,- \tilde{
u}^{c}_{iL})$ & 0\\\hline
\quad$\hat{L}_{a}\sim({\bf1},{\bf2},{\bf1},-1)$ & $(\nu_{a},\,l_{a})_{aL}$
& $\frac{1}{2}$ & $(\tilde{ \nu}_{aL},\,\tilde{ l}_{aL})$ & 0\\
\quad$\hat{L}^{c}_{a}\sim({\bf1},{\bf1},{\bf2},1)$ & $(l^{c}_{a},\,-
\nu^{c}_{a})_{aL}$ & $\frac{1}{2}$ & $(\tilde{ l}^{c}_{aL},\,- \tilde{ \nu
}^{c}_{aL})$ & 0\\\hline
\quad$\hat{\Delta}_{L}\sim({\bf1},{\bf3},{\bf1},2)$ & $\left(
\begin{array}
[c]{cc}%
\frac{\delta_{L}^{+}}{\sqrt{2}} & \delta_{L}^{++}\\
\delta_{L}^{0} & \frac{-\delta_{L}^{+}}{\sqrt{2}}
\end{array}
\right) $ & 0 & $\left(
\begin{array}
[c]{cc}%
\frac{\tilde{\delta}_{L}^{+}}{\sqrt{2}} & \tilde{\delta}_{L}^{++}\\
\tilde{\delta}_{L}^{0} & \frac{-\tilde{\delta}_{L}^{+}}{\sqrt{2}}%
\end{array}
\right) $ & $\frac{1}{2}$\\
\quad$\hat{\Delta}^{\prime}_{L}\sim({\bf1},{\bf3},{\bf1},-2)$ & $\left(
\begin{array}
[c]{cc}%
\frac{\delta_{L}^{\prime-}}{\sqrt{2}} & \delta_{L}^{\prime0}\\
\delta_{L}^{\prime--} & \frac{-\delta_{L}^{\prime-}}{\sqrt{2}}%
\end{array}
\right) $ & 0 & $\left(
\begin{array}
[c]{cc}%
\frac{\tilde{\delta}_{L}^{\prime-}}{\sqrt{2}} & \tilde{\delta}_{L}^{\prime0}\\
\tilde{\delta}_{L}^{\prime--} & \frac{-\tilde{\delta}_{L}^{\prime-}}{\sqrt{2}}%
\end{array}
\right) $ & $\frac{1}{2}$\\
\quad$\hat{\delta}^{c}_{L}\sim({\bf1},{\bf1},{\bf3},-2)$ & $\left(
\begin{array}
[c]{cc}%
\frac{\lambda_{L}^{-}}{\sqrt{2}} & \lambda_{L}^{0}\\
\lambda_{L}^{--} & \frac{-\lambda_{L}^{-}}{\sqrt{2}}%
\end{array}
\right) $ & 0 & $\left(
\begin{array}
[c]{cc}%
\frac{\tilde{\lambda}_{L}^{-}}{\sqrt{2}} & \tilde{\lambda}_{L}^{0}\\
\tilde{\lambda}_{L}^{--} & \frac{-\tilde{\lambda}_{L}^{-}}{\sqrt{2}}%
\end{array}
\right) $ & $\frac{1}{2}$\\
\quad$\hat{\delta}^{\prime c}_{L}\sim({\bf1},{\bf1},{\bf3},2)$ & $\left(
\begin{array}
[c]{cc}%
\frac{\lambda_{L}^{\prime+}}{\sqrt{2}} & \lambda_{L}^{\prime++}\\
\lambda_{L}^{\prime0} & \frac{-\lambda_{L}^{\prime+}}{\sqrt{2}}%
\end{array}
\right) $ & 0 & $\left(
\begin{array}
[c]{cc}%
\frac{\tilde{\lambda}_{L}^{\prime+}}{\sqrt{2}} & \tilde{\lambda}_{L}%
^{\prime++}\\
\tilde{\lambda}_{L}^{\prime0} & \frac{-\tilde{\lambda}_{L}^{\prime+}}{\sqrt
{2}}%
\end{array}
\right) $ & $\frac{1}{2}$\\\hline
\quad$\hat{\Phi} \sim\left(  {\bf1},{\bf2},{\bf2},0\right) $ & $\left(
\begin{array}
[c]{cc}%
\phi_{1}^{0} & \phi_{1}^{+}\\
\phi_{2}^{-} & \phi_{2}^{0}%
\end{array}
\right) $ & 0 & $\left(
\begin{array}
[c]{cc}%
\tilde{\phi}_{1}^{0} & \tilde{\phi}_{1}^{+}\\
\tilde{\phi}_{2}^{-} & \tilde{\phi}_{2}^{0}%
\end{array}
\right) $ & $\frac{1}{2}$\\
\quad$\hat{\Phi}^{\prime} \sim\left(  {\bf1},{\bf2},{\bf2},0\right) $ & $\left(
\begin{array}
[c]{cc}%
\chi_{1}^{0} & \chi_{1}^{+}\\
\chi_{2}^{-} & \chi_{2}^{0}%
\end{array}
\right) $ & 0 & $\left(
\begin{array}
[c]{cc}%
\tilde{\chi}_{1}^{0} & \tilde{\chi}_{1}^{+}\\
\tilde{\chi}_{2}^{-} & \tilde{\chi}_{2}^{0}%
\end{array}
\right) $ & $\frac{1}{2}$\\\hline
\end{tabular}
\caption{Particle content of SUSYLRT.}
\label{tab:SUSYLRT}
\end{table}

The particle content of the model is given at Tab.(\ref{tab:SUSYLRT}) (for
recent work see for example \cite{phenosusylr} and references therein). In 
parentheses it appears the transformation properties under the respective 
$(SU(3)_{C},SU(2)_{L},SU(2)_{R},U(1)_{B-L})$. The
Lagrangian is given at Ref.~\cite{cmmc}.

The Left-Right models may have doubly charged Scalars, see Tab.(\ref{tab:SUSYLRT}), as 
a consequence of this when we construct their 
supersymmetric version, we get double charged charginos \cite{huitu}.

The most general superpotential $W$ \cite{susylr} is given by
\begin{eqnarray}
W  & =&M_{\Delta}Tr(\hat{\Delta}_{L}\hat{\Delta}_{L}^{\prime})+M_{\delta^{c}
}Tr(\hat{\delta}_{L}^{c}\hat{\delta}_{L}^{\prime c})+\mu_{1}Tr(\imath\tau
_{2}\hat{\Phi}\imath\tau_{2}\hat{\Phi})+\mu_{2}Tr(\imath\tau_{2}\hat{\Phi
}^{\prime}\imath\tau_{2}\hat{\Phi}^{\prime})\nonumber\\
& +&\mu_{3}Tr(\imath\tau_{2}\hat{\Phi}\imath\tau_{2}\hat{\Phi}^{\prime}
)+f_{ab}Tr(\hat{L}_{a}\imath\tau_{2}\hat{\Delta}_{L}\hat{L}_{b})+f_{ab}
^{c}Tr(\hat{L}_{a}^{c}\imath\tau_{2}\hat{\delta}_{L}^{c}\hat{L}_{b}
^{c})\nonumber\\
& +&h_{ab}^{l}Tr(\hat{L}_{a}\hat{\Phi}\imath\tau_{2}\hat{L}_{b}^{c})+\tilde
{h}_{ab}^{l}Tr(\hat{L}_{a}\hat{\Phi}^{\prime}\imath\tau_{2}\hat{L}_{b}
^{c})+h_{ij}^{q}Tr(\hat{Q}_{i}\hat{\Phi}\imath\tau_{2}\hat{Q}_{j}
^{c})\nonumber\\
& +&\tilde{h}_{ij}^{q}Tr(\hat{Q}_{i}\hat{\Phi}^{\prime}\imath\tau_{2}\hat
{Q}_{j}^{c})+W_{NR}.
\label{suplr}
\end{eqnarray}
Where $h^{l},\tilde{h}^{l},h^{q}$ and $\tilde{h}^{q}$ are the Yukawa couplings
for the leptons and quarks, respectively, and $f$ and $f^{c}$ are the
couplings for the triplets scalar bosons. We must emphasize that due to the
conservation of $B-L$ symmetry, $\Delta_{L}^{\prime}$ and 
$\delta_{L}^{\prime c}$ do not couple with the leptons and quarks. Here $W_{NR}$ denotes (possible)
non-renormalizable terms arising from higher scale physics or Planck scale
effects \cite{Chacko:1997cm}. This model can be embedded in a supersymmetric
grand unified theory as $SO(10)$ \cite{moha}.

Applying the invariance conditions under $R$-parity transformation in 
Eq.(\ref{suplr}) we get the following equations 
\begin{eqnarray}
n_{\Delta_{L}}+n_{\Delta^{\prime}_{L}}&=&0,
n_{\delta^{c}_{L}}+n_{\delta^{\prime c}_{L}}=0,
2n_{\Phi}=2n_{\Phi^{\prime}}=0,
n_{\Phi}+n_{\Phi^{\prime}}=0, 
2n_{L}+n_{\Delta_{L}}=0,
2n_{L^{c}}+n_{\delta^{c}_{L}}=0, \nonumber \\
n_{L}+n_{L^{c}}+n_{\Phi}&=&0,
n_{L}+n_{L^{c}}+n_{\Phi^{\prime}}=0,
n_{Q}+n_{Q^{c}}+n_{\Phi}=0,
n_{Q}+n_{Q^{c}}+n_{\Phi^{\prime}}=0.
\label{1d}. 
\end{eqnarray}
For example, choosing 
\begin{eqnarray} 
n_{\Delta_{L}}&=&-1, n_{\Delta^{\prime}_{L}}=1,
n_{\delta^{c}_{L}}=1, n_{\delta^{\prime c}_{L}}=-1,
n_{\Phi}=n_{\Phi^{\prime}}=0, \nonumber \\
n_{L}&=& \frac{1}{2},  n_{L^{c}}=- \frac{1}{2},
n_{Q}= \frac{1}{2},  n_{Q^{c}}=- \frac{1}{2},
\label{sym1} 
\end{eqnarray} 
the chiral superfields of this model will transform as 
\begin{eqnarray}
\hat{\Delta}_{L}(x,\theta,\bar{\theta}) 
&\stackrel{R_{d}}{\longmapsto}&
\hat{\Delta}_{L}(x,\theta,\bar{\theta}),
\hat{\Delta}^{\prime}_{L}(x,\theta,\bar{\theta}) 
\stackrel{R_{d}}{\longmapsto} 
\hat{\Delta}^{\prime}_{L}(x,\theta,\bar{\theta}),
\hat{\delta}^{c}_{L}(x,\theta,\bar{\theta}) 
\stackrel{R_{d}}{\longmapsto} 
\hat{\delta}^{c}_{L}(x,\theta,\bar{\theta}),
\hat{\delta}^{\prime c}_{L}(x,\theta,\bar{\theta}) 
\stackrel{R_{d}}{\longmapsto} 
\hat{\delta}^{\prime c}_{L}(x,\theta,\bar{\theta}), \nonumber \\
\hat{\Phi}(x,\theta,\bar{\theta}) 
&\stackrel{R_{d}}{\longmapsto}& 
\hat{\Phi}(x,\theta,\bar{\theta}),
\hat{\Phi}^{\prime}(x,\theta,\bar{\theta}) 
\stackrel{R_{d}}{\longmapsto} 
\hat{\Phi}^{\prime}(x,\theta,\bar{\theta}), 
\hat{L}(x,\theta,\bar{\theta}) 
\stackrel{R_{d}}{\longmapsto}
- \hat{L}(x,\theta,\bar{\theta}),
\hat{L}^{c}(x,\theta,\bar{\theta}) 
\stackrel{R_{d}}{\longmapsto} 
- \hat{L}^{c}(x,\theta,\bar{\theta}), \nonumber \\
\hat{Q}(x,\theta,\bar{\theta}) 
&\stackrel{R_{d}}{\longmapsto}&
- \hat{Q}(x,\theta,\bar{\theta}),
\hat{Q}^{c}(x,\theta,\bar{\theta}) 
\stackrel{R_{d}}{\longmapsto}
- \hat{Q}^{c}(x,\theta,\bar{\theta}),
\label{Rpa1c} 
\end{eqnarray} 
In terms of the field components, we obtain 
\begin{equation}
\begin{array}{rcl}
 H(x) &\stackrel{R_{d}}{\longmapsto}& H(x), \\
 \tilde{H}(x) &\stackrel{R_{d}}{\longmapsto}&
  - \tilde{H}(x),
\end{array}
\begin{array}{rcl}
 \tilde{f}(x) &\stackrel{R_{d}}{\longmapsto}& - \tilde {f}(x), \\
 \Psi(x) &\stackrel{R_{d}}{\longmapsto}&  \Psi(x).
\end{array}
\label{Rpa2} 
\end{equation}
where $H$ is a usual scalar, $\tilde{H}$ is the higgsinos, while 
$\Psi$ is the fermion and $\tilde{f}$ is a sfermion.

Equation (\ref{Rpa2}) suggests to classify the particles into two
types of so called R-even and R-odd. Here the R-even particles
$(R_d=+1)$ include the quarks, the leptons and the Higgs bosons. Whereas, the
R-odd particles $(R_d=-1)$ are their superpartners, i.e., 
neutralinos, charginos, squarks and sleptons. Therefore,
R-parity is parity of R-charge of the continuous $\mathrm{U}(1)$
R-symmetry and defined by 
\begin{equation} \label{eq:rp01}
\hbox{R-parity}=\left\{
\begin{array}{l}
+1 \hspace*{0.5cm} \hbox{for ordinary particles,} \vspace{2mm} \\
-1 \hspace*{0.5cm} \hbox{for their superpartners.}
\end{array}  \right.
\end{equation}

The above intimate connection between R-parity and baryon number,
lepton number conservation laws can be made explicitly by
re-expressing (\ref{eq:rp01}) in terms of the spin $S$ and the
matter-parity $(-1)^{3(B-L)}$ as follows \cite{farrar78}:
\begin{equation}
\label{eq:rp02} 
\hbox{R-parity} = (-1)^{2S} (-1)^{3(B-L)}.
\end{equation}
Therefore, all scalar fields $(S=0)$ can be assigned $R$ values
\begin{itemize} 
\item Usual scalars: $B=L=0 \Longrightarrow R=+1$, 
\item Sleptons: $B=0,\ L=1 \Longrightarrow R=-1$, 
\item Squarks: $B= \frac{1}{3},\ L=0 \Longrightarrow R=-1$,
\end{itemize} 
Analogously for fermions $(S=1/2)$ 
\begin{itemize}
\item Higgsinos $B=0,\ L=0 \Longrightarrow R=-1$, 
\item Leptons: $B=0,\ L=1 \Longrightarrow R=+1 $,
\item Quarks: $B= \frac{1}{3},\ L=0 \Longrightarrow R=+1$.
\end{itemize}

To finish this section, let us note that there will be a lot of
other choices of the charges due to the action of R-symmetry which 
in a general way can be written as \cite{rdiscreta}
\begin{eqnarray} 
\Phi \longrightarrow
e^{2in_{\Phi}\frac{2\pi}{N}}\Phi.
\end{eqnarray} 
Here it is similar to a
$Z_N$ symmetry. Among those choices, there is a possibility which is 
known as Lepton Parity where we choose
\begin{eqnarray}
n_{\Delta_{L}}&=&0, n_{\Delta^{\prime}_{L}}=0,
n_{\delta^{c}_{L}}=0, n_{\delta^{\prime c}_{L}}=0,
n_{\Phi}=n_{\Phi^{\prime}}=0, \nonumber \\
n_{L}&=& \frac{1}{2},  n_{L^{c}}=- \frac{1}{2},
n_{Q}=0,  n_{Q^{c}}=0.
\label{lparity}
\end{eqnarray}
The reason to this name is due the fact that we get the following
transformation of the superfields
\begin{equation} 
( \hat{L}, \hat{L}^{c}) \to - ( \hat{L}, \hat{L}^{c}). \label{pabar}
\end{equation} 
while all others fields are even. All the terms in the superpotential, 
see Eq.(\ref{suplr}), are allowed by the Lepton Parity. It is important to say that in this case the formula $(-1)^{3(B-L)}$ does not hold on this case. 

Other, interesting, possibility 
is the Baryon Parity. On this case we want this transformation
\begin{equation} 
( \hat{Q}, \hat{Q}^{c}) \to - ( \hat{Q}, \hat{Q}^{c}). \label{pabar1}
\end{equation} 
while all others fields are even. To get this transformation law, we need 
to choose this charges
\begin{eqnarray} 
n_{\Delta_{L}}&=&0, n_{\Delta^{\prime}_{L}}=0,
n_{\delta^{c}_{L}}=0, n_{\delta^{\prime c}_{L}}=0,
n_{\Phi}=n_{\Phi^{\prime}}=0, \nonumber \\
n_{L}&=&0,  n_{L^{c}}=0,
n_{Q}= \frac{1}{2},  n_{Q^{c}}=- \frac{1}{2}.
\label{bparity}
\end{eqnarray}
These Baryon Parity doesn't eliminate any terms in our superpotential. As in the previous case, we can not write the $R$ charge as $(-1)^{3(B-L)}$.

In the case of the MSSM we choose the Lepton Parity, defined at 
Eq.(\ref{lparity}), and the Baryon Parity, see Eq.(\ref{bparity}), in order to avoid 
the proton decay. It happens because using these ``new" parity we can 
eliminate some terms in the superpotential of the MSSM \cite{hall,banks,rv1}. 

However it is not the case on this model. The only difference between the traditional 
$R$-parity and these parities defined above is that now the Eq.(\ref{eq:rp02}) is not 
satified.

\section{SUSYLRD}

\begin{table}[t]
\center
\renewcommand{\arraystretch}{1.5}
\begin{tabular}
[c]{|l|cc|cc|}\hline
Superfield & Usual Particle & Spin & Superpartner & Spin\\\hline\hline
\quad$\hat{V}^{\prime}$ (U(1)) & $B_{m}$ & 1 & $\tilde{B}\,\,$ & $\frac{1}{2}%
$\\
\quad$\hat{V}^{i}_{L}$ ($SU(2)_{L}$) & $W^{i}_{mL}$ & 1 & $\tilde{W}_{L}^{i}$
& $\frac{1}{2}$\\
\quad$\hat{V}^{i}_{R}$ ($SU(2)_{R}$) & $W^{i}_{mR}$ & 1 & $\tilde{W}_{R}^{i}$
& $\frac{1}{2}$\\
\quad$\hat{V}^{a}_{c} (SU(3))$ & $g^{a}_{m}$ & 1 & $\tilde{ g}^{a}$ &
$\frac{1}{2}$\\\hline
\quad$\hat{Q}_{i}\sim({\bf3},{\bf2},{\bf1},1/3)$ & $(u_{i},\,d_{i})_{iL}$ &
$\frac{1}{2}$ & $(\tilde{ u}_{iL},\,\tilde{ d}_{iL})$ & 0\\
\quad$\hat{Q}^{c}_{i}\sim({\bf3^{*}},{\bf1},{\bf2},-1/3)$ & $(d^{c}_{i}%
,\,-u^{c}_{i})_{iL}$ & $\frac{1}{2}$ & $(\tilde{ d}^{c}_{iL},\,- \tilde{
u}^{c}_{iL})$ & 0\\\hline
\quad$\hat{L}_{a}\sim({\bf1},{\bf2},{\bf1},-1)$ & $(\nu_{a},\,l_{a})_{aL}$
& $\frac{1}{2}$ & $(\tilde{ \nu}_{aL},\,\tilde{ l}_{aL})$ & 0\\
\quad$\hat{L}^{c}_{a}\sim({\bf1},{\bf1},{\bf2},1)$ & $(l^{c}_{a},\,-
\nu^{c}_{a})_{aL}$ & $\frac{1}{2}$ & $(\tilde{ l}^{c}_{aL},\,- \tilde{ \nu
}^{c}_{aL})$ & 0\\\hline
\quad$\hat{\chi}_{1L} \sim\left(  {\bf1},{\bf2},{\bf1},1 \right) $ & $(\chi_{1L}^{+},
\,\ \chi_{1L}^{0})$ & 0 & $(\tilde{\chi}_{1L}^{+}, \,\ \tilde{\chi}_{1L}^{0})$
& $\frac{1}{2}$\\
\quad$\hat{\chi}_{2L} \sim\left(  {\bf1},{\bf2},{\bf1},-1 \right) $ & $(\chi_{2L}^{0},
\,\ \chi_{2L}^{-})$ & 0 & $(\tilde{\chi}_{2L}^{0}, \,\ \tilde{\chi}_{2L}^{-})$
& $\frac{1}{2}$\\
\quad$\hat{\chi}^{c}_{3L} \sim\left(  {\bf1},{\bf1},{\bf2},-1 \right) $ & $(\chi_{3L}^{0},
\,\ \chi_{3L}^{-})$ & 0 & $(\tilde{\chi}_{3L}^{0}, \,\ \tilde{\chi}_{3L}^{-})$
& $\frac{1}{2}$\\
\quad$\hat{\chi}^{c}_{4L} \sim\left(  {\bf1},{\bf1},{\bf2},1 \right) $ & $(\chi_{4L}^{+},
\,\ \chi_{4L}^{0})$ & 0 & $(\tilde{\chi}_{4L}^{+}, \,\ \tilde{\chi}_{4L}^{0})$
& $\frac{1}{2}$\\\hline
\quad$\hat{\Phi} \sim\left(  {\bf1},{\bf2},{\bf2},0\right) $ & $\left(
\begin{array}
[c]{cc}%
\phi_{1}^{0} & \phi_{1}^{+}\\
\phi_{2}^{-} & \phi_{2}^{0}%
\end{array}
\right) $ & 0 & $\left(
\begin{array}
[c]{cc}%
\tilde{\phi}_{1}^{0} & \tilde{\phi}_{1}^{+}\\
\tilde{\phi}_{2}^{-} & \tilde{\phi}_{2}^{0}%
\end{array}
\right) $ & $\frac{1}{2}$\\
\quad$\hat{\Phi}^{\prime} \sim\left(  {\bf1},{\bf2},{\bf2},0\right) $ & $\left(
\begin{array}
[c]{cc}%
\chi_{1}^{0} & \chi_{1}^{+}\\
\chi_{2}^{-} & \chi_{2}^{0}%
\end{array}
\right) $ & 0 & $\left(
\begin{array}
[c]{cc}%
\tilde{\chi}_{1}^{0} & \tilde{\chi}_{1}^{+}\\
\tilde{\chi}_{2}^{-} & \tilde{\chi}_{2}^{0}%
\end{array}
\right) $ & $\frac{1}{2}$\\\hline
\end{tabular}
\caption{Particle content of SUSYLRD.}
\label{tab:SUSYLRD}
\end{table}
This model contains the particle content given at
Tab.(\ref{tab:SUSYLRD}). The Lagrangian of this model is given at Ref.~\cite{cmmc}.

The most general superpotential and soft supersymmetry breaking Lagrangian for
this model are:
\begin{eqnarray}
W  & =&M_{\chi}\hat{\chi}_{1}\hat{\chi}_{2}+M_{\chi^{c}}\hat{\chi}_{3}^{c}%
\hat{\chi}_{4}^{c}+\mu_{1}Tr(\tau_{2}\hat{\Phi}\tau_{2}\hat{\Phi})+\mu
_{2}Tr(\tau_{2}\hat{\Phi}^{\prime}\tau_{2}\hat{\Phi}^{\prime})+\mu_{3}%
Tr(\tau_{2}\hat{\Phi}\tau_{2}\hat{\Phi}^{\prime})\nonumber\\
& +&h_{ab}^{l}Tr(\hat{L}_{a}\hat{\Phi}\imath\tau_{2}\hat{L}_{b}^{c})+\tilde
{h}_{ab}^{l}Tr(\hat{L}_{a}\hat{\Phi}^{\prime}\imath\tau_{2}\hat{L}_{b}%
^{c})+h_{ij}^{q}Tr(\hat{Q}_{i}\hat{\Phi}\imath\tau_{2}\hat{Q}_{j}^{c})
\nonumber \\
&+&\tilde{h}_{ij}^{q}Tr(\hat{Q}_{i}\hat{\Phi}^{\prime}\imath\tau_{2}\hat{Q}%
_{j}^{c})
+W_{NR}.\label{suplrd}
\end{eqnarray}

Applying the invariance conditions under $R$-parity transformation in 
Eq.(\ref{suplrd}) we get the following equations 
\begin{eqnarray}
n_{\chi_{1}}+n_{\chi_{2}}&=&0, 
n_{\chi^{c}_{3}}+n_{\chi^{c}_{4}}=0,
2n_{\Phi}=2n_{\Phi^{\prime}}=0,
n_{\Phi}+n_{\Phi^{\prime}}=0, 
2n_{L}+n_{\Delta_{L}}=0,
2n_{L^{c}}+n_{\delta^{c}_{L}}=0, \nonumber \\
n_{L}+n_{L^{c}}+n_{\Phi}&=&0,
n_{L}+n_{L^{c}}+n_{\Phi^{\prime}}=0,
n_{Q}+n_{Q^{c}}+n_{\Phi}=0,
n_{Q}+n_{Q^{c}}+n_{\Phi^{\prime}}=0.
\label{1ddd}. 
\end{eqnarray}
For example, choosing 
\begin{eqnarray} 
n_{\chi_{1}}&=&0, n_{\chi_{2}}=0,
n_{\chi^{c}_{3}}=0, n_{\chi^{c}_{4}}=0,
n_{\Phi}=n_{\Phi^{\prime}}=0, \nonumber \\
n_{L}&=& \frac{1}{2},  n_{L^{c}}=- \frac{1}{2},
n_{Q}= \frac{1}{2},  n_{Q^{c}}=- \frac{1}{2},
\label{sym1dd} 
\end{eqnarray} 
the chiral superfields of this model will transform as 
\begin{eqnarray}
\hat{\chi}_{1}(x,\theta,\bar{\theta}) 
&\stackrel{R_{d}}{\longmapsto}&
\hat{\chi}_{1}(x,\theta,\bar{\theta}),
\hat{\chi}_{2}(x,\theta,\bar{\theta}) 
\stackrel{R_{d}}{\longmapsto}
\hat{\chi}_{2}(x,\theta,\bar{\theta}),
\hat{\chi}^{c}_{3}(x,\theta,\bar{\theta}) 
\stackrel{R_{d}}{\longmapsto}
\hat{\chi}^{c}_{3}(x,\theta,\bar{\theta}),
\hat{\chi}^{c}_{4}(x,\theta,\bar{\theta}) 
\stackrel{R_{d}}{\longmapsto}
\hat{\chi}^{c}_{4}(x,\theta,\bar{\theta}), \nonumber \\
\label{Rpa1cdd} 
\end{eqnarray} 

With this charges we reproduce Eqs.(\ref{Rpa2}, \ref{eq:rp02}).

As happened on the first model analyzed above, if we define 
the $L$-parity and the $B$-parity we can not again eliminate any term in the superpotential defined at Eq.(\ref{suplrd}) and the same comments make at that time are still hold on this case.

\section{Conclusion}

We have showed explicity that the superpotential of the models SUSYLRT and SUSYLRD 
they satisfy the $R$-parity, and in this case its values are given by 
$R = (-1)^{3 (B - L) + 2 S}$. In contrast with the MSSM case on this kind of model 
the $R$-parity does not eliminate any terms in the superpotential.

We show also that they also satisfy the lepton and baryon parity. On this case we can not 
use the expression $(-1)^{3(B-L)}$ to definy the $R$-parity.

\section*{Acknowledgments}
M.C. Rodriguez is supported by Conselho Nacional de Ci\^{e}ncia e 
Tecnologia (CNPq) under the contract number 309564/2006-9.


\begin{thebibliography}{199}

\bibitem{superk} Y. Fukuda {\it et al.} (Super-Kamiokande
Collaboration), Phys. Rev. Lett. {\bf 81}, 1562 (1998); {\bf 82},
2644 (1999); {\bf 85}, 3999 (2000); Y. Suzuki, Nucl. Phys. B,
Proc. Suppl. {\bf 77}, 35 (1999); Y. Fukuda {\it et al.}
(Super-Kamiokande Collaboration), Phys. Rev. Lett. {\bf 81}, 1158
(1998); {\bf 82}, 1810 (1999); S. Fukuda {\it et al.,} Phys. Rev.
Lett. {\bf 86}, 5651 (2001).

\bibitem{hanoiconf} H. Ts. Wong,  in {\it the Proceedings of
Hanoi Forum on Frontiers of Basic Science: Towards New Physics:
Earth and Space Science, Mathematics}, Hanoi, Vietnam, 27-29 Sep
2005, Osaka University Press (2006), pp. 74 - 79,
[arXiv:hep-ex/0510001].

\bibitem{fogli} G. L. Fogli, E. Lisi, A. Marrone and A.
Palazzo, [arXiv:hep-ph/0506083].

\bibitem{seesaw}
P. Minkowski, Phys. lett. {\bf B67 }, 421 (1977); M.~Gell-Mann,
P.~Ramond, and R.~Slansky, \emph{Supergravity} (P.~van Nieuwenhuizen
et al. eds.), North Holland, Amsterdam, 1980, p.~315; T.~Yanagida,
in \emph{Proceedings of the Workshop on the Unified Theory and the
Baryon Number in the Universe} (O.~Sawada and A.~Sugamoto, eds.),
KEK, Tsukuba, Japan, 1979, p.~95; S.~L. Glashow, \emph{The future of
elementary particle physics}, in
  \emph{Proceedings of the 1979 Carg{\`e}se Summer Institute on
Quarks and Leptons} (M.~L{\'e}vy et al. eds.), Plenum Press, New
York, 1980, pp.~687; R.~N. Mohapatra and G.~Senjanovi{\'c}, Phys.
Rev. Lett. \textbf{44}, 912 (1980).

\bibitem{rad} A. Zee, Phys. Lett. B93, 389 (1980); B161,
141 (1985); Nucl. Phys. B264, 99 (1986); L. Wolfenstein, Nucl. Phys.
B175, 93 (1980); K. S. Babu, Phys. Lett. B 203, 132 (1988); D.
Chang, W.-Y. Keung, and P. B. Pal, Phys. Rev. Lett. 61, 2420 (1988);
J. T. Peltonieri, A. Yu. Smirnov, and J. W. Valle, Phys. Lett. B
286, 321 (1992); D. Choodhury, R. Gandhi, J. A. Gracey, and B.
Mukhopadhyaya, Phys. Rev. D50, 3468 (1994); D. Chang and A. Zee,
Phys. Rev. D 61 (2000) 071303 (R); C. Jarlskog, M. Matsuda, S.
Skalhauge and M. Tanimoto, Phys. Lett. B 449 (1999) 240; M. Frigerio
and A. Yu. Smirnov,  JHEP 0302:004, (2003).

\bibitem{ps74} J.C. Pati and A. Salam, Phys. Rev. {\bf D10}, 275 (1974); 
R.N. Mohapatra and J.C. Pati, {\it ibid} {\bf D11}, 566; 2558 (1975); 
G. Senjanovi\'{c} and R.N. Mohapatra, {\it ibid} {\bf D12}, 1502 (1975).
For details see G. Senjanovi\'c, Nucl. Phys. {\bf B153}, 334 (1979). 

\bibitem{bt78}M.A.B. Beg and H.S. Tsao,  Phys. Rev. Lett. {\bf 41} (1978) 278; 
R.N. Mohapatra and G. Senjanovi\'c, Phys. Lett. {\bf 79B} (1978) 28.

\bibitem {susylr}K. Huitu, J. Maalampi and M. Raidal, {\sl Nucl. Phys.}{\bf B420}, 
449 (1994); C.S.Aulakh,A.Melfo and G.Senjanovic,
{\sl Phys.Rev.}{\bf D57},4174 (1998); G. Barenboim and N. Rius,
{\sl Phys. Rev.}{\bf D58}, 065010, (1998); N. Setzer and S. Spinner,
{\sl Phys. Rev.} {\bf D71}, 115010 (2005).

\bibitem {doublet}K. S. Babu.B. Dutta and R.N. Mohapatra, {\sl Phys.Rev.}
{\bf D65}:016005, (2002).

\bibitem{m86} R. N. Mohapatra,  Phys. Rev. {\bf D34} 3457 (1986);
A. Font, L. E. Ib\'a\~nez and F. Quevedo,  Phys. Lett. {\bf B228} 79 (1989); 
L. Ib\'a\~nez and G. Ross,  Phys. Lett. {\bf B260} 291 (1991); 
S. P. Martin,  Phys. Rev. {\bf D 46} 2769 (1992).


\bibitem{seesaw1}M. Gell-Mann, P. Ramond and R. Slansky, in {\it Supergravity}, eds. P. van Niewenhuizen and
D.Z. Freedman (North Holland 1979); 
T. Yanagida, in Proceedings of {\it Workshop on Unified Theory and Baryon number in the Universe}, eds.
O. Sawada and A. Sugamoto (KEK 1979);
R. N. Mohapatra and G. Senjanovi{\'c}, Phys. Rev. Lett. {\bf 44} (1980) 912.

\bibitem {mfrank}M. Frank, {\sl Phys. Lett.} {\bf 540},269 (2002).

\bibitem{r1} A. Salam and J. Strathdee, Nucl. Phys. B {\bf 87}, 85 (1975).

\bibitem{r2} P. Fayet, Nucl. Phys. B {\bf 90}, 104 (1975).

\bibitem{barbier} R. Barbier {\it et al.}, Phys. Rept. {\bf 420},
1 (2005), [arXiv:hep-ph/0406039].

\bibitem{moreau} G. Moreau, [arXiv:hep-ph/0012156].

\bibitem{hung} P.V. Dong, D.T. Huong, M.C. Rodriguez and H.N. Long,  {\sl Eur.Phys.J.} {\bf C48}, 229 (2006).

\bibitem{cmmc} C.M. Maekawa and M. C. Rodriguez, JHEP \textbf{04}(2006),031. 

\bibitem{huitu}K.Huitu, J.Maalampi, M. Raidal, {\sl Nucl. Phys.}{\bf B420}, 449 (1994);
G. Barenboim, K. Huitu, J. Maalampi and M. Raidal, {\it Phys.Lett.}{\bf B394},132 (1997);
K.Huitu, J.Maalampi, A. Pietil\"{a} and M. Raidal, 
{\it Nucl.Phys.}{\bf B487}, 27 (1997); 
M. Franck, {\sl Phys. Rev.}{\bf D62}, 053004-1 (2000); M. C. Rodriguez, arXiv:0706.3065 [hep-ph].

\bibitem {Chacko:1997cm}Z.~Chacko and R.~N.~Mohapatra, {\sl Phys. Rev.}
{\bf D58}, 015003 (1998); B.~Dutta and R.~N.~Mohapatra, {\sl Phys. Rev.}
{\bf D59}, 015018 (1999).

\bibitem {moha}R. N. Mohapatra, hep-ph/9801235. 

\end{thebibliography}
\end{document}